\documentclass{appolb}
\usepackage{epsfig}

\begin{document}
\title{ HIGHLY RELATIVISTIC MOTIONS \\ OF SPINNING PARTICLES\\ ACCORDING TO MATHISSON EQUATIONS %
}
\author{Roman Plyatsko
\address Pidstryhach Institute for Applied Problems\\ in Mechanics and
Mathematics \\ Ukrainian National Academy of Sciences\\ Naukova
3-b, 79060 Lviv, Ukraine}

\maketitle
\begin{abstract}
The physical effects following from the Mathisson equations at the
highly relativistic motions of a spinning test particle relative
to a Schwarz\-schild mass are discussed. The corresponding
numerical estimates are presented.

\end{abstract}

\section{Introduction}

 During 70 years the Mathisson equations \cite{1} have being investigated
 by many authors with different intensity. The veri fruitful years
 were 1970s [2--11]. There is interesting remark in \cite{5},
 p.111:
 "The simple act of endowing a black hole with angular momentum has
led to an unexpected richness of possible physical phenomena. It
seems appropriate to ask whether endowing the test body with
intrinsic spin might not also lead to surprises". In this context
the question of importance is: can spin of a test particle
essentially change its world line and trajectory? To answer this
question it is useful to consider the Mathisson equations both in
their traditional form and in the terms of the local (tetrad)
quantities connected with the moving particle. The initial form of
the Mathisson equations is \cite{1}
\begin{equation}\label{1}
\frac D {ds} \left(mu^\lambda + u_\mu\frac {DS^{\lambda\mu}}
{ds}\right)= -\frac {1} {2} u^\pi S^{\rho\sigma}
R^{\lambda}_{\pi\rho\sigma},
\end{equation}
\begin{equation}\label{2}
\frac {DS^{\mu\nu}} {ds} + u^\mu u_\sigma \frac {DS^{\nu\sigma}}
{ds} - u^\nu u_\sigma \frac {DS^{\mu\sigma}} {ds} = 0,
\end{equation}
where $u^\lambda$ is the 4-velocity of a spinning particle,
$S^{\mu\nu}$ is the antisymmetric tensor of spin, $m$ and $D/ds$
are, respectively, the mass and the covariant derivative with
respect to the proper time $s$; $R^{\lambda}_{\pi\rho\sigma}$ is
the Riemann curvature tensor of the spacetime. (Greek indices run
1, 2, 3, 4 and Latin indices 1, 2, 3.) Equations (\ref{1}),
(\ref{2}) were supplemented by the condition \cite{1}
\begin{equation}\label{3}
S^{\mu\nu} u_\nu = 0,
\end{equation}
which first was used in electrodynamics \cite{12}. Later the
condition
\begin{equation}\label{4}
S^{\mu\nu} P_\nu = 0
\end{equation}
was introduced \cite{2, 13}, where
\begin{equation}\label{5}
P^\nu = mu^\nu + u_\mu\frac {DS^{\nu\mu}}{ds}.
\end{equation}
Concerning the physical meaning of conditions (\ref{3}) and
(\ref{4}) see, {\it e.g.,} \cite{14}.

Besides $S^{\mu\nu}$, the 4-vector of spin $s_\lambda$ is also
used in the literature, where by definition \cite{3}
\begin{equation}\label{6}
 s_\lambda=\frac{1}{2}\sqrt{-g}\varepsilon_{\lambda\mu\nu\sigma}u^\mu
 S^{\nu\sigma}
\end{equation}
($g$ is the determinant of the metric tensor) with the relation

\begin{equation}\label{7}
s_\lambda s^\lambda=\frac12S_{\mu\nu}S^{\mu\nu}=S_0^2,
\end{equation}
where $S_0$ is the constant of spin.

\section{Mathisson equations in representation \\ of Ricci's coefficients of rotation}

For transformation of equations (\ref{1}), (\ref{2}) under
condition (\ref{3}) we use the relations for the comoving
orthogonal tetrads $\lambda_\mu^{(\nu)}$:
\begin{equation}\label{8}
dx^{(i)} = \lambda_\mu^{(i)}dx^\mu = 0,\quad
 dx^{(4)} = \lambda_\mu^{(4)}dx^\mu = ds
\end{equation}
(here indices of the tetrad are placed in the parentheses). For
convenience, we choose the first local coordinate axis as oriented
along the spin, then
$$
s_{(1)} \neq 0,\quad s_{(2)} = 0,
$$
\begin{equation}\label{9}
 s_{(3)}=0,\quad s_{(4)}=0,
\end{equation}
and $|s_{(1)}|=S_0$.

By (\ref{3}), (\ref{8}), (\ref{9}) it follows from (\ref{2}) that
$\gamma_{(i)(k)(4)}=0$, {\it i.e.,} the known condition for the
Fermi transport, where $\gamma_{(\alpha)(\beta)(\gamma)}$ are
Ricci's coefficients of rotation. From equations (1) one can find
\begin{equation}
\label{10} m\gamma_{(1)(4)(4)} + {s}_{(1)} R_{(1)(4)(2)(3)} = 0,
\end{equation}
\begin{equation}
\label{11}
 m\gamma_{(2)(4)(4)} + {s}_{(1)}(R_{(2)(4)(2)(3)}
 - \dot \gamma_{(3)(4)(4)}) = 0,
\end{equation}
\begin{equation}
\label{12} m\gamma_{(3)(4)(4)} + {s}_{(1)}(R_{(3)(4)(2)(3)} + \dot
\gamma_{(2)(4)(4)}) = 0,
\end{equation}
where a dot denote the usual derivatives with respect to $s$. The
Ricci coefficients of rotation $\gamma_{(i)(4)(4)}$ have the
physical meaning as the components $a_{(i)}$ of the 3-acceleration
of a spinning particle relative to geodesic free fall as measured
by the comoving observer.

In the linear in spin approximation equations
(\ref{10})--(\ref{12}) can be written as \cite{15}
\begin{equation}\label{13}
\gamma_{(i)(4)(4)}\equiv a_{(i)} =-{s_{(1)}\over
m}R_{(i)(4)(2)(3)}.
\end{equation}
It is known that within the framework of this approximation the
physical consequences following from equations (\ref{1}),
(\ref{2}) coincide under conditions (\ref{3}) and (\ref{4})
\cite{16, 17}.

By definition of the gravitoelectric $E_{(k)}^{(i)}$ and the
gravitomagnetic $B_{(k)}^{(i)}$ components we have \cite{18}
\begin{equation}\label{14}
E_{(k)}^{(i)}=R^{(i)(4)}_{}{}{}{}{}{}_{(k)(4)},
\end{equation}
\begin{equation}\label{15}
B_{(k)}^{(i)}=-\frac12 R^{(i)(4)}_{}{}{}{}{}{}_{(m)(n)}
\varepsilon^{(m)(n)}_{}{}{}{}{}{}_{(k)}.
\end{equation}
So, according to (\ref{13}), (\ref{15}) the acceleration $a_{(i)}$
is determined by $B_{(k)}^{(i)}$.

\section{Case of a Schwarzschild metric}

Let us consider equation (\ref{13}) for Schwarzschild metric in
standard coordinates $x^1=r,\quad x^2=\theta,\quad
x^3=\varphi,\quad x^4=t$. The motion of an observer relative to
Schwarzschild's mass $M$ can be described by the orthonormal frame
$\lambda_\mu^{(\nu)}$. For expediency and without loss of
generality we assume that the first tetrad axis is perpendicular
to plane determined by the direction of observer motion and the
radial direction ($\theta=\pi/2$), and the second  axis coincides
with the direction of motion. Then the non-zero components of
$B_{(k)}^{(i)}$ are \cite{19}:
$$
B^{(1)}_{(2)}=B^{(2)}_{(1)}= \frac{3Mu_\parallel u_\perp}
{r^3\sqrt{u_4u^4-1}}\left(1-\frac{2M}{r}\right)^{-1/2},
$$
\begin{equation}\label{16}
B^{(1)}_{(3)}=B^{(3)}_{(1)}= \frac{3M u_\perp^2 u^4}
{r^3\sqrt{u_4u^4-1}}\left(1-\frac{2M}{r}\right)^{1/2},
\end{equation}
where $u_\parallel=\dot r$ is the radial component of the observer
4-velocity , $u_\perp=r\dot\varphi$ is the tangential component.

Let a spinning particle be comoving with the observer and its spin
is oriented along the first tetrad axis. By (\ref{15}), (\ref{16})
relation (\ref{13}) can be written as
\begin{equation}\label{17}
a_{(i)}=\frac{s_{(1)}}{m}B^{(1)}_{(i)}.
\end{equation}
From (\ref{16}), (\ref{17}) it is easy to see that the
acceleration $a_{(i)}$ is not equal to $0$ if and only if $u_\perp
\ne 0$, {\it i.e.,} is caused by the gravitational spin-orbit
interaction. (The gravitational spin-orbit and spin-spin
interactions in the post-Newtonian approximation were investigated
in \cite{4}.) By (\ref{16}), (\ref{17}) we have
\begin{equation}\label{18}
\vert \vec a_{s.-o.}\vert = {M\over{r^2}}{3s_{(1)}\vert u_\perp
\vert\over{mr}} \sqrt{1+u^2_\perp },
\end{equation}
where $\vert \vec a_{s.-o.}\vert \equiv \sqrt{a_{(1)}^2 +
a_{(2)}^2 + a_{(3)}^2}$ is the absolute value of the gravitational
spin-orbit acceleration. While investigating possible effects of
spin on the particle's motion it is necessary to take into account
the condition for a spinning test particle \cite{4}
\begin{equation}\label{19}
\varepsilon\equiv\frac{|s_{(1)}|}{mr} \ll 1.
\end{equation}
According to (\ref{18}), (\ref{19}), the two limiting cases are
essentially different in their physical consequences: (1) at low
velocity, when ($|u_\perp|\ll 1$), we have $ \vert \vec
a_{s.-o.}\vert \ll {M/{r^2}},$ where $M/{r^2}$ is the Newtonian
value of the free fall acceleration; (2) at high velocity, when
$|u_\perp|\gg 1,$ for any small $\varepsilon$ we can indicate such
sufficiently large value $|u_\perp|$ for which the value $\vert
\vec a_{s.-o.}\vert$ is of order $M/{r^2}.$ That is, in the second
case the motion of a spinning particle essentially differs from
the geodesic motion \cite{15}. We stress that this conclusion is
obtained from the point of view of the comoving observer. Is the
similar conclusion valid for an observer which, for example, is at
rest relative to the Schwarzschild mass? To answer this question
it is necessary to investigate the corresponding solutions of
equations (\ref{1}), (\ref{2}).

The interesting partial solutions of equations (\ref{1}),
(\ref{2}) in a Schwarzschild spacetime were studied in \cite{19,
20}. Namely, it is shown that the circular highly relativistic
orbits of a spinning particle exist in the small neighborhood of
the value $r=3M$ (both for $r>3M$ and $r<3M$) with
\begin{equation}\label{20}
 |u_\perp|=\frac{3^{1/4}}{\sqrt{\varepsilon}}
\end{equation}
({\it i.e.,} according to (\ref{19}), (\ref{20}) we have
$u_\perp^2\gg 1$). These orbits differ from the highly
relativistic geodesic circular orbits, which exist only for
$r>3M$. Besides, it is known that a particle without spin and with
non-zero mass of any velocity close to the velocity of light,
starting in the tangential direction from the position $r=3M$,
will fall on Schwarzschild's horizon within a finite proper time,
whereas a spinning particle will remain indefinitely on the
circular orbit $r=3M$ due to the interaction of its spin with the
gravitational field, {\it i.e.,} this interaction compensates the
usual (geodesic) attraction.

The highly relativistic circular orbits determined by (\ref{20})
are practically common for equations (\ref{1}), (\ref{2}) at
conditions (\ref{3}) and (\ref{4}) \cite{20}. Outside the small
neighborhood of $r=3M$, for $2M<r<3M$, equations (\ref{1}),
(\ref{2}) admit circular highly relativistic orbits as well,
however, only under condition (\ref{3}) \cite{14, 20}. Similarly
to (\ref{20}), the value $|u_\perp|$ on these orbits is of order
$1/\sqrt{\varepsilon}$.

Some non-circular essentially non-geodesic orbits with the initial
values of $|u_\perp|$ of order $1/\sqrt{\varepsilon}$ were
computed in \cite{14}.

\section{Conclusions and numerical estimates }

So, if the tangential component of the particle's velocity is of
order $1/\sqrt{\varepsilon}$, its spin can essentially deviate the
particle's trajectory from the geodesic line, both from the point
of view of the comoving observer and from the point of view of an
observer which is at rest relative to the Schwarzschild mass. (By
(\ref{18}), if $|u_\perp|$ is of order $1/\sqrt{\varepsilon}$, the
acceleration $\vert \vec a_{s.-o.}\vert$ is of order $M/{r^2}$.)

The effect of the considerable space separation of spinning and
non-spinning  particles  takes place for the short time: less than
one or two revolutions of the spinning particles around a
Schwarzschild mass the difference of the radial coordinates
$\Delta r$ becomes comparable with the initial radial coordinate
\cite{14}.

The existence of highly relativistic circular orbits of a spinning
particle in a Schwarzschild field, which differ from geodesic
circular orbits, perhaps, can be discovered in the synchrotron
radiation of protons or electrons.

 Let us estimate the value $\varepsilon=|S_0|/Mm$ for protons
and electrons  in the cases when the Schwarzschild source is 1) a
black hole of mass that is equal to three of the Sun's mass, and
2) a massive black hole of mass that is equal to $10^8$ of the
Sun's mass. In the first case, taking into account the numerical
values of $S_0, M, m$ in the system used, we have for protons and
electrons correspondingly $\varepsilon_p \approx 2\cdot
10^{-20},\quad \varepsilon_e \approx 4\cdot 10^{-17}$, whereas in
the second case $\varepsilon_p \approx 7\cdot 10^{-28},\quad
\varepsilon_e \approx 10^{-24}$. Because for the motion on above
considered orbits the spinning particles must possess the velocity
corresponding  to relativistic Lorentz $\gamma$-factor of order
$1/\sqrt{\varepsilon}$, in the first case we obtain $\gamma_p
\approx 7\cdot 10^9,\quad \gamma_e \approx 2\cdot 10^8$, and in
the second case $\gamma_p \approx 4\cdot10^{13}$,\quad $\gamma_e
\approx 10^{12}$.

As we see, in the case of a massive black hole the necessary
values of $\gamma_p$ and $\gamma_e$ are too high even for
extremely relativistic particles from the cosmic rays. Whereas if
the Schwarzschild source is a black hole of mass that is equal to
3 of the Sun's mass, some particles may move on the circular
orbits considered above. Analysis of a concrete model, closer to
the reality, remains to be carried out. Here we point out that by
the known general relationships for the electromagnetic
synchrotron radiation in the case of protons or electrons on the
considered circular orbits we obtain the values from the gamma-ray
range.

The results of investigation of the highly relativistic
considerable non-geodesic motions of a spinning test particle
according to the Mathisson equations stimulate the analysis of the
corresponding quantum states which are described by the Dirac
equations \cite{19}.

\end{document}